# Structural, electronic, thermal and mechanical properties of $C_{60}$-based fullerene two-dimensional networks explored by first-principles and machine learning

Bohayra Mortazavi*,[a,b]

[a]Department of Mathematics and Physics, Leibniz Universität Hannover,
Appelstraße 11,30167 Hannover, Germany.
[b]Cluster of Excellence PhoenixD (Photonics, Optics, and Engineering–Innovation Across Disciplines), Gottfried Wilhelm Leibniz Universität Hannover, Hannover, Germany.

## Abstract

Recent experimental reports on the realizations of two-dimensional (2D) networks of the $C_{60}$-based fullerenes with anisotropic and nanoporous lattices represent a significant advance, and create exciting prospects for the development of a new class of nanomaterials. In this work, we employed theoretical calculations to explore novel $C_{60}$-based fullerene lattices and subsequently evaluate their stability and key physical properties. After the energy minimization of extensive structures, we could detect novel 2D, 1D and porous carbon $C_{60}$-based networks, with close energies to that of the isolated $C_{60}$ cage. Density functional theory results confirm that the $C_{60}$-based networks can exhibit remarkable thermal stability, and depending on their atomic structure show metallic, semimetallic or semiconducting electronic nature. Using the machine learning interatomic potentials, thermal and mechanical responses of the predicted nanoporous 2D lattices were investigated. The estimated thermal conductivity of the quasi-hexagonal-phase of $C_{60}$ fullerene is shown to be in an excellent agreement with the experimental measurements. Despite of different atomic structures, the anisotropic room temperature lattice thermal conductivity of the fullerene nanosheets are estimated to be in the order of 10 W/mK. Unlike the majority of carbon-based 2D materials, $C_{60}$-based counterparts noticeably are predicted to show positive thermal expansion coefficients. Porous carbon $C_{60}$-based networks are found to exhibit superior mechanical properties, with tensile strengths and elastic modulus reaching extraordinary values of 50 and 300 GPa, respectively. The theoretical results presented in this work provide a comprehensive vision on the structural, energetic, electronic, thermal and mechanical properties of the $C_{60}$-based fullerene networks.

## 1. Introduction

Fullerenes are a class of closed-cage molecules entirely composed of carbon atoms. Classical fullerenes are constructed from only hexagonal and pentagonal carbon rings. Full-carbon fullerenes can exist in diverse forms [1–5], consisting of different number of atoms with various topology of the hexagonal and pentagonal rings, making them among the most extensive molecular systems. The $C_{60}$ [2] is the first fullerene molecule that was experimentally realized in 1994. Fullerenes are known for their interesting structural features and appealing physical and chemical properties, which make them promising candidates for a wide range of applications [6–8]. For example, fullerene cages can exhibit decent electrical conductivity, larger than that of the copper [9], very appealing for applications in electronics and energy storage systems [7]. Furthermore, fullerenes have been found to exhibit biological activities, including antioxidant and antimicrobial properties, which are attractive for pharmaceutical applications [6,8].

The recent report [10] on the synthesis of the quasi-hexagonal-phase of the $C_{60}$ fullerene (qHPC$_{60}$) in the two-dimensional (2D) form, is undoubtedly among the most exciting latest advances in the continuously expanding field of 2D nanomaterials. According to theoretical findings, qHPC$_{60}$ monolayer can show remarkable strength and stability [11–16], low thermal conductivity [5,11] and semiconducting electronic nature [14,15,17,18]. Following the first experimental report by Hou *et al.* [10], two other experimental groups have also recently succeeded in the fabrication of the $C_{60}$ fullerene 2D networks [19,20]. Particularly, Pan *et al.* [19] synthesized new types of the $C_{60}$ fullerene 2D networks, with long-range ordered porous carbon like structures, different from the original qHPC$_{60}$ structure. Aforementioned experimental accomplishments [10,19,20], not only propose innovative synthesis strategies to form diverse fullerene-based 2D networks with nonporous and light-weight structures, but can also be beneficial to explore topology dependent physics and chemistry of carbon nanomaterials.

Inspired by the recent experimental accomplishments [10,19,20], herein density functional theory (DFT) and machine learning based calculations [21–23] were carried out, in order to explore $C_{60}$-based networks and subsequently evaluate their various physical properties. By performing the energy minimization of over 1000 randomly generated atomic configurations, we could detect novel 2D, 1D and porous-carbon $C_{60}$-based networks, with close energies to that of the native $C_{60}$ molecule. Although during the ab-initio molecular dynamics (AIMD) simulations at 1000 K, the majority of predicted structures were found to transform to the pristine $C_{60}$ cage or 1D chains, we could successfully detect numerous thermally, dynamically and mechanically stable $C_{60}$-based networks. DFT calculations reveal that the predicted structures can show diverse electronic properties, ranging from purely metallic to semimetallic and semiconducting. The phonon dispersion relations, phonons' group velocity, thermal expansion coefficient, lattice thermal conductivity and mechanical/failure properties of the predicted $C_{60}$-based networks were elaborately studied using machine learning interatomic potentials (MLIPs) [24–28]. This study not only introduces novel stable $C_{60}$-based networks, but also provides a very comprehensive and useful vision on the structural, energetic, electronic, thermal and mechanical properties of this exciting class of nanomaterials.

## 2. Computational methods

Vienna ab-initio simulation package (VASP) [29,30] was used to perform DFT calculations by employing Perdew-Burke-Ernzerhof (PBE) and generalized gradient approximation (GGA) methods, Grimme's DFT-D3 [31] van der Waals (vdW) dispersion correction and a 500 eV kinetic energy cutoff. Optimization of lattice parameters and atomic positions were conducted on the basis of the conjugate gradient method using a 3×3×1 Monkhorst-Pack [32] k-point grid, with energy and force convergence criteria of $10^{-5}$ eV and 0.002 eV/Å, respectively. In order to prevent vdW interactions along the thickness, a large vacuum spacing (more than 14 Å) was adjusted. Ab-initio molecular dynamics (AIMD) simulations were carried out at 1000 K with a time step of 1 fs for around 8 ps, to inspect the thermal stability. VESTA [33] and OVITO [34] packages were used to illustrate the atomic structures and VASPKIT package [35] was also employed to find symmetry points of the Brillouin zone (BZ) for the analysis of band structures. We developed moment tensor potentials (MTPs) [36] to investigate the phononic, thermal and mechanical properties, employing the MLIP package [37]. To create the training dataset, AIMD calculations were performed over unitcells with a time step of 1 fs, DFT-D3 dispersion correction, and a coarser 2×2×1 grid. We considered unstrained and biaxially strained systems (-5 to +15% strain) in the AIMD calculations under varying temperatures from 1 to 1500 K, in order to capture occurring atomic environments during the simulations, using the same approach as that of our recent studies [5,38–40]. 14000, 12000 and 8000 configurations were generated for the 2D, porous carbon and 1D $C_{60}$-based structures, respectively (the entire AIMD datasets are given in the data availability section). The complete AIMD datasets were then subsampled and around 600 configurations were selected for the fitting of the preliminary MTPs. The accuracy of the first MTPs were evaluated over the full AIMD datasets, and configurations with poor extrapolation grades [41] were identified and added to the original dataset, which was then used to train the second and final MTPs with a cutoff distance of 3.5 Å (trained MTPs are also provided in the data availability section). The outstanding accuracy and computational efficiency of the aforementioned approach for examining the thermal and mechanical properties have been confirmed in our recent studies [38–40]. The phonon dispersion relations were obtained based on the developed MTPs [42], over 5×5×1 supercells and by using the small displacement method of the PHONOPY package [43].

The LAMMPS package [44] was employed to examine the thermal and mechanical properties based on the developed MTPs, using a fixed time step of 0.5 fs. We conducted classical non-equilibrium molecular dynamics (NEMD) simulations to evaluate the length effect on the thermal transport of the $C_{60}$-based 2D networks. In the NEMD method, first the systems were equilibrated using the Nosé-Hoover barostat and thermostat method (NPT). Next, a few rows of atoms at boundary of the models along the heat transfer direction were fixed, and a 20 K temperature difference was applied between the two end of the model using the Nosé-Hoover thermostat (NVT) method, while the remaining part of the system was simulated under the constant energy (NVE) ensemble. The lattice thermal conductivity was calculated using the 1D Fourier's law, based on the applied heat flux by the NVT and averaged temperature gradient along the sample's length [45,46]. The complex mechanical and deformation responses were explored using the quasi-static uniaxial tensile loading, which can accurately reproduce the DFT results at the ground state [38–40]. In the aforementioned approach, the loading was applied with a fixed strain step of 0.0025, applied in every 100000 time increments, while the structural relaxation was accomplished with the NPT method to ensure uniaxial stress condition. The stress values in the second half of the NPT relaxation were averaged to obtain smooth stress-strain relations.

## 3. Results and discussions

We first describe the developed strategy for the reconstruction of $C_{60}$-based fullerene networks. In this approach, after the energy minimization of the isolated $C_{60}$ cage, it was used to generate randomly connected 2D structures. First, three random rotations were applied along the three Cartesian directions with respect to the center of the atomic mass of the $C_{60}$ molecule. The purpose of these rotations is to ensure that the orientation of the molecule is completely arbitrary, and that is not aligned in any particular direction. In the next step, the simulation box dimensions along the planar directions were randomly altered to form preliminary bonds with the periodic images. This step is necessary to ensure that the resulting lattice is periodic and can be extended infinitely in the 2D form. The generated lattice was then subjected to a DFT energy minimization step of the cell dimensions and atomic positions in order to find the minimum energy configuration. The calculations were conducted for more than 1000 randomly fabricated lattices, and subsequently the total energy of the systems were calculated. By comparing the energies of different structures, the lattices with the lowest energies were identified. For the majority of cases, the randomly generated 2D structures nonetheless converted either to the original 0D cage or 1D chains.

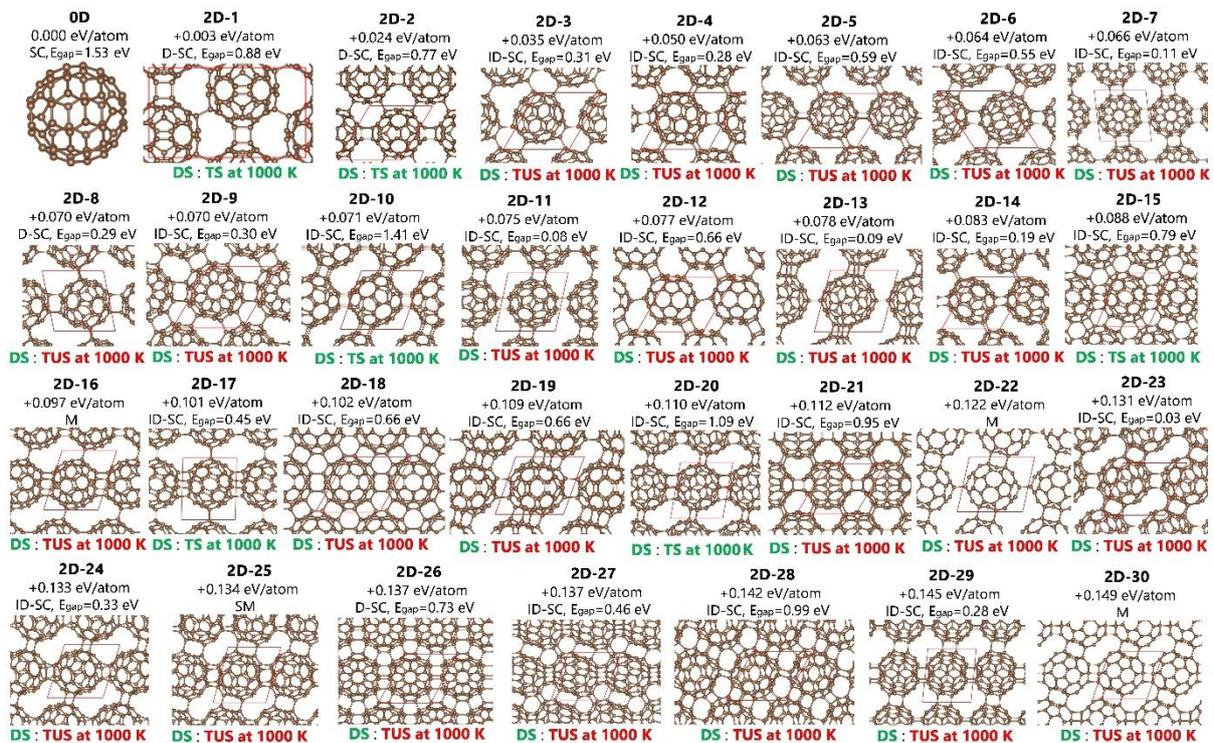

**Fig. 1**, $C_{60}$ cage and the $C_{60}$-based 2D crystalline networks, ranked according to their total energy (value under the structure name) with respect to that of the isolated $C_{60}$ molecule, 2D-1 to 2D-30. The 2D-1 lattice represents the experimentally fabricated qHPC$_{60}$ monolayer. D-SC, ID-SC and E$_{gap}$, represent the direct, indirect and the corresponding PBE/GGA band gap value for the semiconductors, respectively, and SM and M indicate semimetallic and metallic nature, respectively. DS stands for the dynamical stability, TS and TUS at 1000 K represent thermal stability and instability after AIMD simulations at 1000 K, respectively.

By conducting the DFT calculations, we could successfully predict novel full-C$_{60}$ 2D structures, and the first 29 lowest structures were selected, as shown in Fig. 1. In this illustration, we also show the $C_{60}$ molecule (0D lattice) and the experimentally fabricated qHPC$_{60}$ monolayer (2D-1 lattice), alongside the 29 lowest energy structures generated in this study. For the simplification of referring to these structures, we rank them according to their total energy with respect to that of the $C_{60}$ molecule, from 2D-1 to 2D-30. Interestingly, while 2D-2, 2D-4 and 2D-12 lattices exhibit similar nanopore morphologies as that of the experimentally realized qHPC$_{60}$, their atomic structures are not exactly identical and that is why they show slightly different energies. From the first glance, the 2D-1 and 2D-2 lattices seems completely identical, however, in the 2D-1 lattice the orientation of the second $C_{60}$ cage in the unitcell is flipped with respect to the first one. In this work, since only a single $C_{60}$ molecule was considered in the modeling, we could not explore the complete possible atomic configurations and that explains why the original qHPC$_{60}$ lattice was not reconstructed. Therefore, in order to detect more extensive stable structures, it is necessary to include more than one randomly oriented $C_{60}$ molecules in the primitive cell, which will become computationally exceedingly expensive, when employing the DFT method for the geometry optimization. The substitution of the DFT with MLIPs can be a highly appealing alternative to resolve the aforementioned bottleneck, which can be implemented in the oncoming studies. Moreover, by conducting the calculations for more random configurations, certainly other low energy systems could be also detected. As an important observation, although the bonding architectures are different for the generated systems, their total energies are very close. For example, the 2D-30 lattice exhibits a total energy, only 0.146 eV/atom less negative than that of the experimentally fabricated qHPC$_{60}$. As an example, in our recent studies we could detect BCN [28] and BC$_2$N [45] 2D lattices that are by 0.166 and 0.306 eV/atom, more energetically stable than those of the experimentally realized BCN [47] and BC$_2$N [48] nanosheets, respectively. It is thus clear that due to the very close energies of the $C_{60}$-based 2D networks, various atomic configurations may co-exist in the experimental samples. It is moreover noticeable that we were unable to find any symmetrical structure along the two planar perpendicular directions, which indicates the dominance of anisotropic electronic, optical, mechanical and heat transport along the $C_{60}$-based 2D networks. Among the generated structures, 2D-10 is the only lattice with symmetrical bonding architecture between $C_{60}$-cages, made of identical four C-C bonds. We also analyzed the bonding mechanism in the fullerene 2D networks, using the electron localization functions (ELF) [49], which as expected, revealed the pure covalent interactions in the $C_{60}$-based fullerene networks. We also analyzed the dynamical stability of the generated $C_{60}$ 2D networks shown in Fig. 1 (find the data availability section) on the basis of the phonon dispersion relations acquired using the MTP method [42], which reveal that the entire acoustic and optical phonon modes are free of imaginary frequencies, confirming the dynamical stability of the generated 2D lattices.

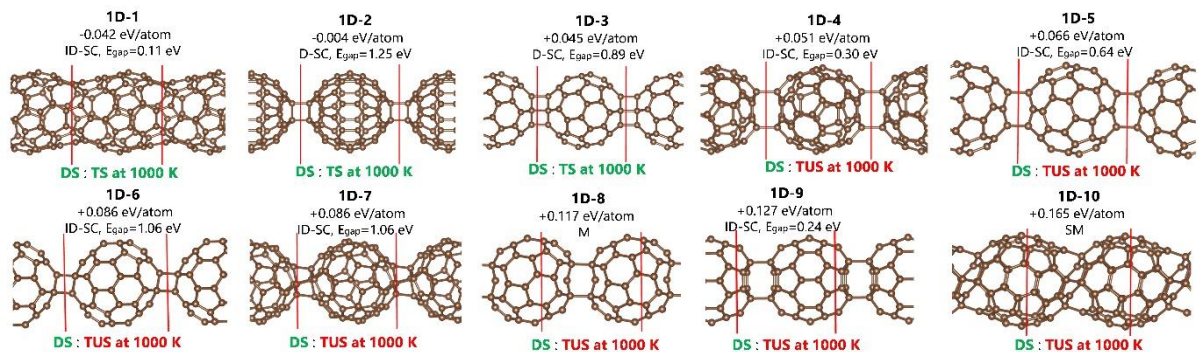

**Fig. 2**, Generated 1D $C_{60}$-based networks, ranked according to their total energy (value under the structure name) with respect to that of the isolated $C_{60}$ molecule, 1D-1 to 1D-10. D-SC, ID-SC and $E_{gap}$, represent the direct, indirect and the corresponding PBE/GGA band gap value for semiconductors, respectively, and SM and M indicate semimetallic and metallic nature, respectively. DS stands for the dynamical stability, TS and TUS at 1000 K represent for the thermal stability and instability after AIMD simulations at 1000 K, respectively.

Although the main objective of the herein study was to investigate the formation of the $C_{60}$-based 2D networks, but during the energy minimization step, remarkable portion of the originally produced 2D structures evolved to 1D $C_{60}$-based chains. We could predict several new $C_{60}$-based 1D structures, which are illustrated in Fig. 2. Worth noting that the formation of the 1D-2 structure was already discussed in the original experimental work by Hou *et al.* [10]. Interestingly, we could detect the formation of the first $C_{60}$-based nanotube (1D-1 structure in Fig. 2), which shows a lower energy than the native $C_{60}$ cage. As it is clear, the positive energies of the 1D-3 to 1D-10 lattices with respect to the original $C_{60}$ cage reveal that they are metastable. Phonon dispersion results confirm the dynamical stability of the generated 1D chains (find the data availability section).

During the energy minimization step of the randomly generated $C_{60}$-based 2D structures, in some cases the formation of 2D structures with nanotube-like hollow structures along one direction or with some carbon atoms pushed inside/outside of the original $C_{60}$ fullerene were observed. For the aforementioned latter case, the free or irregular carbon atoms we removed and the energy minimization step was subsequently re-conducted. In these cases, the generated networks resemble the long-range ordered porous carbon like structures, as those realized most recently by Pan *et al.* [19]. In Fig. 3, the first 26 most energetically stable constructed $C_{60}$-based porous carbon 2D lattices are presented. In consistency with earlier observation for the $C_{60}$-based 2D lattices, total energies are generally close for systems with very different atomic structures. It is noticeable that the least energetically stable lattice presented in Fig. 3 (PC-26) shows a total energy by only around 0.281 eV/atom less negative than that of the isolated $C_{60}$ cage. Among the theoretically produced lattices, only two (PC-2 and PC-20 lattices in Fig. 3) were found to be dynamically unstable, and the phonon dispersion relations of the remaining structures were observed to be completely free of imaginary frequencies, confirming their dynamical stability. AIMD simulations were furthermore carried out at 1000 K for around 8 ps, in order to inspect the thermal stability of the generated lattices. As summarized in Fig. 1, from the total 29 produced novel structures, 24 lattices were found to deform either to 0D or 1D lattices or undergo bond breakages, revealing their thermal instability at 1000 K. As an interesting observation, from the 26 constructed $C_{60}$-based porous carbon 2D lattices, 12 structures were found to be thermally stable at 1000 K. It is nonetheless worth noting that 1000 K is generally considered as a high temperature, which is certainly beyond the ambient condition for the majority of practical applications [50–54]. Owing to the formation of strong covalent interactions in the predicted $C_{60}$-based networks, those predicted unstable at 1000 K, may stay completely intact at lower temperatures. AIMD results clearly indicate that the formation of nanotube-like structures could enhance the thermal stability of the $C_{60}$-based networks. The atomic structures of the produced lattices and the corresponding data for their thermal stability and phonon dispersion relations are entirely given in the data availability section.

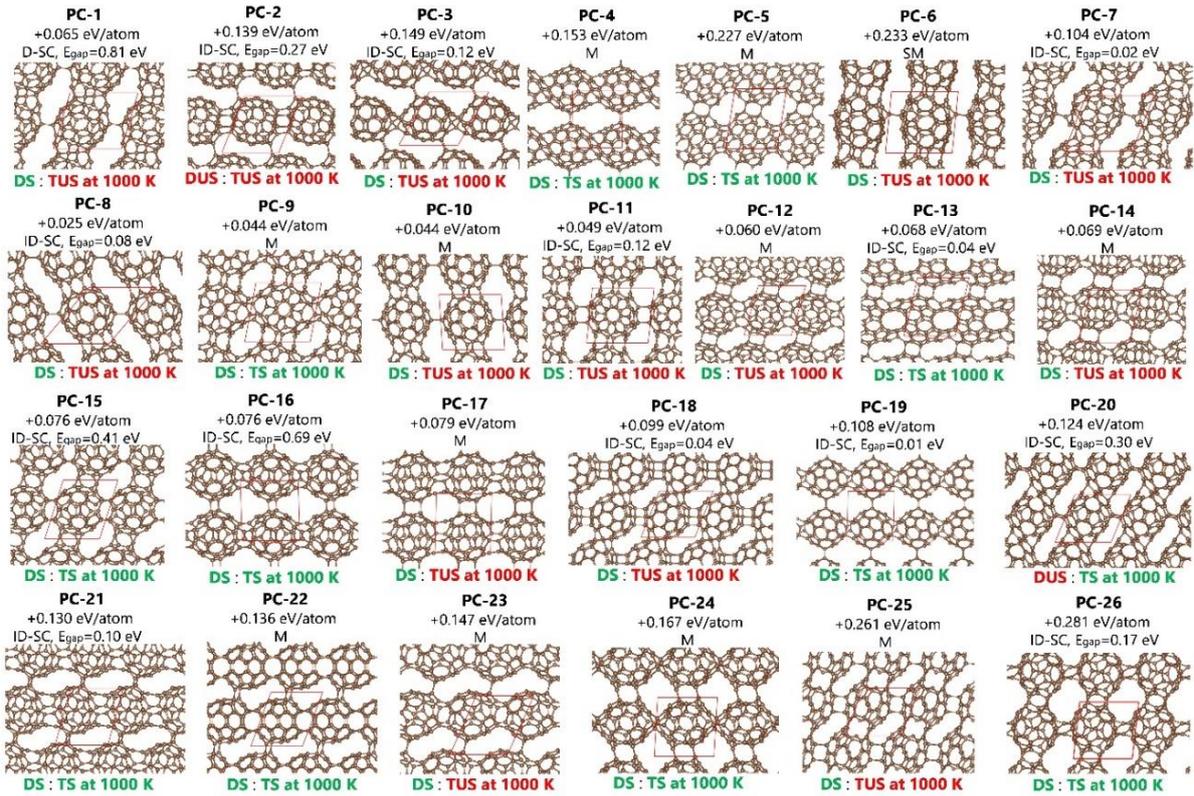

**Fig. 3**, Produced C$_{60}$-based porous carbon (PC) lattice, along with their total energy with respect to that of the isolated C$_{60}$ molecule (value under the structure name). D-SC, ID-SC and E$_{gap}$, represent the direct, indirect and the corresponding PBE/GGA band gap value for semiconductors, respectively, and SM and M indicate semimetallic and metallic nature, respectively. DS and DUS stand for the dynamical stability and instability, respectively, TS and TUS at 1000 K represent for the thermal stability and instability after AIMD simulations at 1000 K, respectively.

After the examination of the structural features and energetic, thermal and dynamical stability of the predicted C$_{60}$-based networks, we next briefly investigate their electronic features on the basis of the PBE/GGA functional. The electronic band gap of the isolated C$_{60}$ molecule was predicted to be 1.53 eV, which underestimates the experimentally measured value of 1.86±0.1 eV [55]. In the present work, the VASPKIT package [35] was employed to find the symmetry points for the analysis of band structures and moreover to summarize the electronic properties. The high-symmetry points of the BZ, corresponding KPOINTS path for the VASP calculations, band Gap value, positions of the valance band maximum and conduction band minimum and complete PBE/GGA electronic band structure for every structure are included in the data availability section. As it is clear, C$_{60}$-based networks depending on the atomic structure can exhibit diverse electronic natures, ranging from metallic to semimetallic and indirect or direct gap semiconducting. In Fig. 4, the electronic band structures for several representative C$_{60}$-based networks are illustrated. It is shown that the 2D-1 and 2D-2 monolayers both are direct gap semiconductors at the Γ point, and noticeably despite their very identical atomic structures and exactly the same bonding architectures between $C_{60}$-cages, the former exhibits a band gap by 0.11 eV larger than that of the latter. The predicted 2D-8 lattice is found to be also a direct gap semiconductor, with a narrow gap of 0.29 eV and relatively sharp dispersions for the both valance and conduction bands, which can be an indication of decent carrier mobilities in this system. For the 2D $C_{60}$-based crystalline networks, the maximum PBE/GGA band gap was found to be around 1.4 eV, which is distinctly higher than that of the 0.81 eV for the porous carbon counterparts. It is worth mentioning that the PBE/GGA method systematically underestimates the band gap value for the semiconducting or insulating systems, and therefore the employment of hybrid functionals like HSE06 [56] or state-of-the-art GW [57] methods are necessary to find more accurate estimations. Generally, the presented results indicate that the metallic electronic nature is more conspicuous in the $C_{60}$-based porous carbons networks.

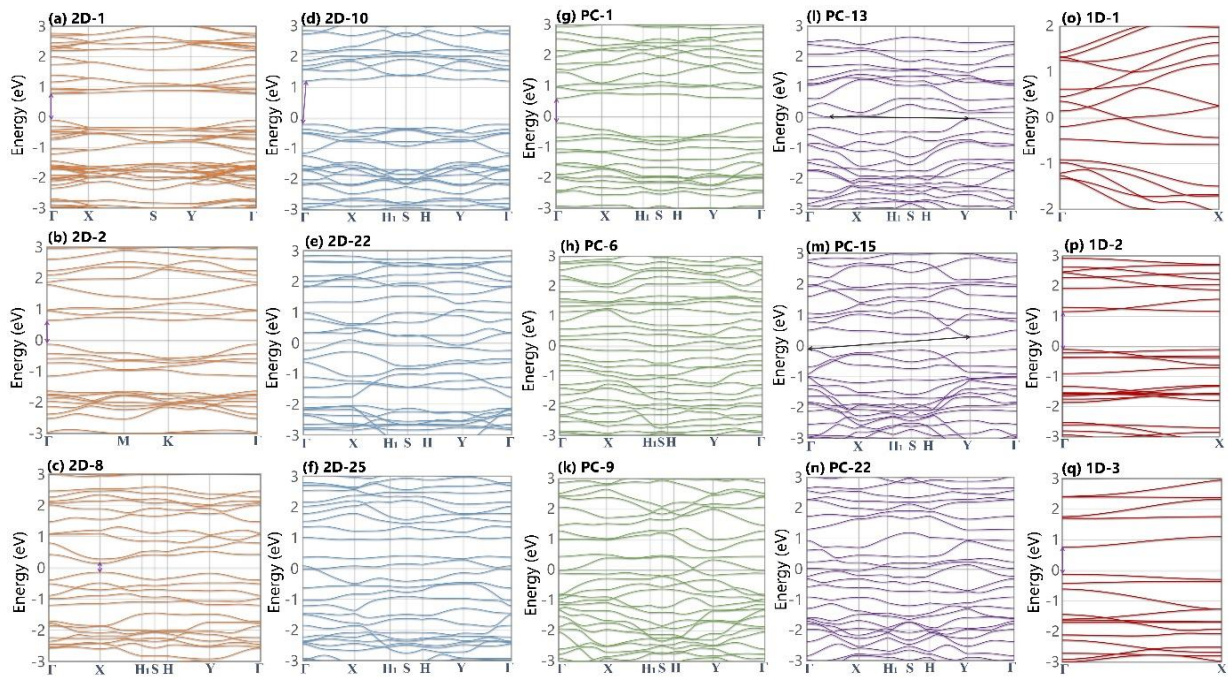

**Fig. 4**, Electronic band structures of the $C_{60}$-based networks on the basis of PBE/GGA functional.

After evaluating the electronic properties, we now shift our attention to the investigation of the thermal properties of the $C_{60}$-based networks. First, we evaluate the lattice thermal conductivity at room temperature by conducting the NEMD simulations on the basis of the trained MTPs. Examining the effect of length on the estimated lattice thermal conductivity is critical in the NEMD simulations of the heat conduction, because of the fact that the fixed atoms at both ends of the model may restrict the contribution of long wavelength acoustic phonons. Because of the excessive computational costs of NEMD calculations for the evaluation of room temperature thermal conductivity, in this section we only consider 2D-1, 2D-10 and PC-21 lattices. The thermal conductivity of the qHPC$_{60}$ (2D-1 lattice) bulk system has been recently examined experimentally by Meirzadeh *et al.* [20], which is thus an excellent choice for evaluating the accuracy of the MTP-based NEMD modeling. The 2D-10 network is also the most symmetrical lattice, and the only one with identical four C-C bonds connecting the C$_{60}$-cages along different directions. The PC-21 porous carbon network also includes a nanotube-like structure in one direction connected by two C-C bonds along the perpendicular direction, very appealing to explore the anisotropic thermal transport in these systems. In Fig. 5, first the phonon dispersion curves of the 2D-1, 2D-10 and PC-21 monolayers based on the developed MTPs are presented. It is observable from the insets of dispersion curves that none of the three acoustic branches exhibit imaginary frequencies, and the longitudinal acoustic phonon modes show widest dispersions in these networks. The corresponding group velocities for the phonon branches for every structure is also calculated and presented in insets of Fig. 5. It is noticeable that the maximum group velocity for the phonons belongs to longitudinal acoustic modes, which is close for different systems and around 11 km/s. While the isotropic character for the phonons group velocity is observable along the symmetrical points of the BZ for the 2D-10 lattice, anisotropic behaviors are conspicuous along the two other considered counterparts. As a consistent behavior, optical phonon modes with frequencies larger than 10 THz generally tend to exhibit rather flat dispersions, with low phonon group velocities, in agreement with the results depicted in Fig. 5 insets. Additionally, the in-plane acoustic and whole optical phonon modes display significant band crossing throughout the entire frequency range, indicating high scattering rates and therefore short phonon lifetimes for the majority of heat carriers in these systems. The combination of low phonon group velocity and high scattering rates implies negligible contribution of optical modes on the heat transport in these nanomaterials. The flexural acoustic phonon mode in these systems shows narrower dispersions and thus lower group velocities than the two in-plane counterparts, but they generally present no intersection with other bands in these systems, revealing their longest lifetime. These preliminary observations suggest low lattice thermal conductivity along the C$_{60}$-based 2D networks.

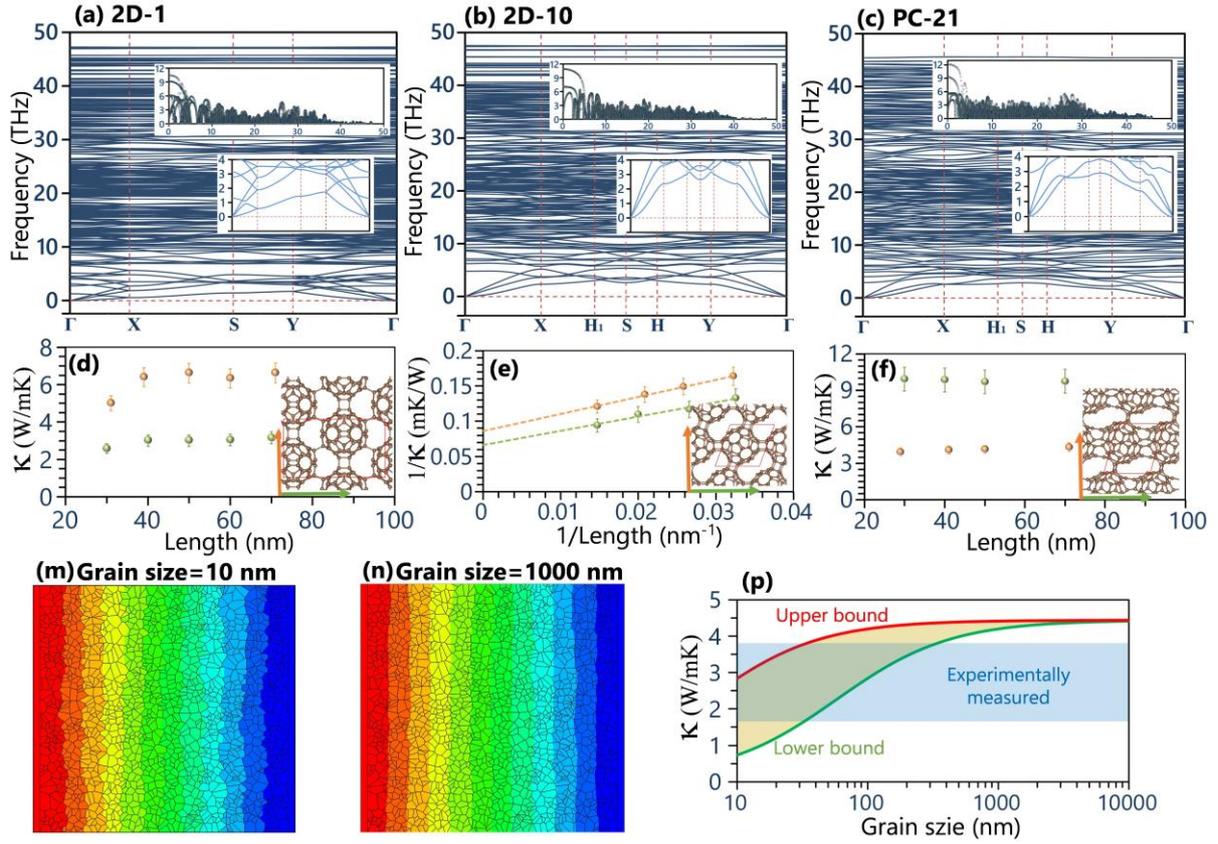

Fig. 5, MTP-based predicted phonon dispersion relation of the (a) 2D-1, (b) 2D-10 and (c) PC-21 $C_{60}$-based nanosheets. In the presented results, upper and lower insets illustrate the phonons' group velocity in the km/s unit and zoomed view of the low frequency modes in the THz unit, respectively. NEMD results for the length and direction effect on the room temperature phononic thermal conductivity of the (d) qHPC$_{60}$, (e) 2D-10 and (f) PC-21 lattices. Finite element results for the temperature distribution of samples with average grain size of (m) 10 nm and (n) 1000 nm (color coding represents the high (red) to low (blue) temperatures). (p) First-principles multiscale model [38,58] for the effective lattice thermal conductivity of the polycrystalline qHPC$_{60}$ at room temperature, for two interfacial thermal conductance values of the 0.3 (upper bound) and 0.03 GW/m$^2$K (lower bound).

Fig. 5 also shows the NEMD results for the effect of length on the lattice thermal conductivity of the considered $C_{60}$-based 2D networks, predicted at room temperature. As it is clear, for the 2D-1 and PC-21 lattices, the thermal conductivity is completely converged with respect to the length. The room temperature lattice thermal conductivity of the 2D-1 and PC-21 $C_{60}$-based monolayers along the X(Y) directions, are predicted to be 3.1±0.3(6.1±0.5) and 9.7±0.8(4.3±0.3) W/mK, respectively. On the other side, for the 2D-10 counterpart, increasing trends in the thermal conductivity by the length is observable. As a routine approach, the diffusive thermal conductivity was obtained by using the inverse relationships and extrapolation of linear fits [59]. The room temperature lattice thermal conductivity of the 2D-10 monolayer along the X(Y) directions are predicted to be 15.2±1.2 (11.6±1.0) W/mK.

Consistent with the conclusion of our recent study on the $C_{36}$ fullerene 2D network [5], the room temperature thermal conductivity of the fullerene-based 2D networks are generally in the order of 10 W/mK. This order is generally rather high to be appealing for thermoelectric energy conversion and also very low for the application in thermal management systems. It is moreover clear that $C_{60}$-based porous carbons networks can exhibit higher thermal conductivity than the crystalline counterparts, due to the facilitated thermal transport along the nanotube-like direction. In order to validate the predicted values, we consider the measured thermal conductivity of the qHPC$_{60}$ (2D-1 lattice) bulk system by Meirzadeh *et al.* [20]. In their measurements because of complex microstructure and technical difficulties, directional thermal conductivity of the single-crystalline lattices was not obtained and instead an averaged value of 2.7±1.0 W/mK over the entire system was reported [20]. Likely to the majority of 2D nanomaterials, the experimentally realized fullerene-based 2D networks may also show polycrystalline nature and therefore the constructed microstructure may affect transport properties. In order to evaluate the effective lattice thermal conductivity of the polycrystalline qHPC$_{60}$ nanosheets, we utilized the finite element simulations with ABAQUS/Standard and python scripting, and developed a model consisting of 1600 individual grains constructed using the Voronoi approach with mirror symmetry [58,60]. The anisotropic thermal conductivity tensor for the individual qHPC$_{60}$ grains was defined by randomly selecting the orientation, using the NEMD results presented earlier. However, it is critical to define the interfacial thermal conductance between connecting grains in the constructed continuum models [58,60]. Direct evaluation of the interfacial thermal conduce between grains with different misorientation angles made of various dislocation cores, using the molecular dynamics simulations, is nonetheless an extensive, complex and computationally expensive task [58]. Interestingly, according to our previous studies on the graphene [60], MoS$_2$ [61] and h-BN [62] nanosheets, despite their very different lattice thermal conductivities, the effective interfacial thermal conductance between connecting grains was estimated to be in the order of $\kappa_{grain}$/(30 nm), according to the diffusive thermal conductivity of the single-crystalline monolayer ($\kappa_{grain}$). As it is clear, one can consider an upper and lower bound for the interfacial thermal conductance in 2D systems, using the $\kappa_{grain}$/(10 nm) and $\kappa_{grain}$/(100 nm), respectively. In this work, on the basis of the predicted thermal conductivity of the single crystalline qHPC60 monolayer (2D-1 lattice) along the X direction ($\kappa_{grain}$=3.1 W/mK), we consequently assumed upper and lower bound values of the 0.3 and 0.03 GW/m²K for the effective interfacial conductance in the polycrystalline qHPC$_{60}$ sheets. Aforementioned values were subsequently used to define the contact thermal property in the finite element calculations. In Fig. 5p, the effective lattice thermal conductivity of the polycrystalline qHPC$_{60}$ for domain sizes ranging from 10 nm to 10 μm are presented for the two considered interfacial thermal conductances, along with the measured range by Meirzadeh *et al.* [20]. As shown in Fig. 5m, for the average grain size of 10 nm, relatively constant temperature distributions are apparent within the grains, and the established temperature profile for the entire sample is mostly due to jumps at the grain boundaries. For the larger average grain size of 1000 nm, the temperature distribution becomes ostensible inside the grains, implying that the effect of the grain boundaries is considerably suppressed as compared with the smaller grain size of 10 nm. From the results shown in Fig. 5p, it is clear that for the both assumed interfacial conductance values, the effective thermal conductivity fully converges for samples with average grain sizes larger than 3 μm. Remarkably, by considering the measured experimental values [20], it is conspicuous that the developed first-principles multiscale model [38,58] could very precisely reproduce the lattice thermal conductivity of the qHPC$_{60}$ system, not only confirming the outstanding accuracy but also revealing the underlying possible mechanisms for large scattering in the experimentally reported value [20].

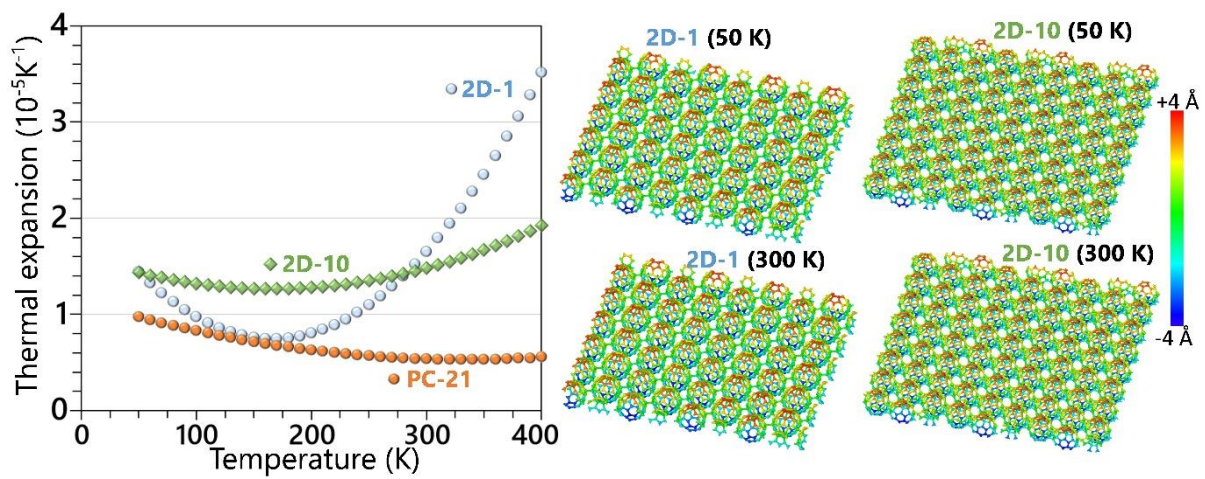

Fig. 6, Predicted thermal expansion coefficient of the single-layer (a) 2D-1, (b) 2D-10 and (c) PC-21 C$_{60}$-based networks as a function of temperature. Snapshots of the 2D-1 and 2D-10 monolayers at 50 and 300 K, in which color coding depict the out-of-plane position of carbon atoms with respect to the center of the mass.

The MTP-based classical models developed in this study were also used to investigate the thermal expansion coefficients of the 2D-1, 2D-10 and PC-21 C$_{60}$-based monolayers, using our previously validated approach [63]. The thermal expansion coefficient as a function of temperature was evaluated using the $\frac{1}{A}\frac{dA}{dT}$ relation, where $A$ is the projected area of the system and $T$ is the temperature [63,64]. By fitting polynomial curves to the projected area and temperature data points, averaged over five independent calculations, the thermal expansion coefficients as a function of temperature [63] were obtained. The results presented in Fig. 6, show that the room temperature thermal expansion coefficient of the single-layer 2D-1, 2D-10 and PC-21 $C_{60}$-based lattices are positive, estimated to be +1.65, +1.48 and +0.54 ×$10^{-5}$ K$^{-1}$, respectively. The thermal expansion coefficient of the single-layer graphene at room temperature was predicted to be −2.95×$10^{-6}$ K$^{-1}$ [63]. As illustrated in the Fig. 6 results for the atomic structures, in the $C_{60}$-based nanosheets as the temperature increases, due to high bending rigidity in these systems the formation of out-of-plane wrinkles is very limited. The bonds elongations in these systems under higher atomic vibrations accordingly appear mostly by the expansion in the projected area. In contrast for the case of graphene and the majority of fully flat nanosheets [46,63], by increasing the temperature the lattice undergoes considerable out-of-plane wrinkles, which results in yielding negative thermal expansion coefficients [63].

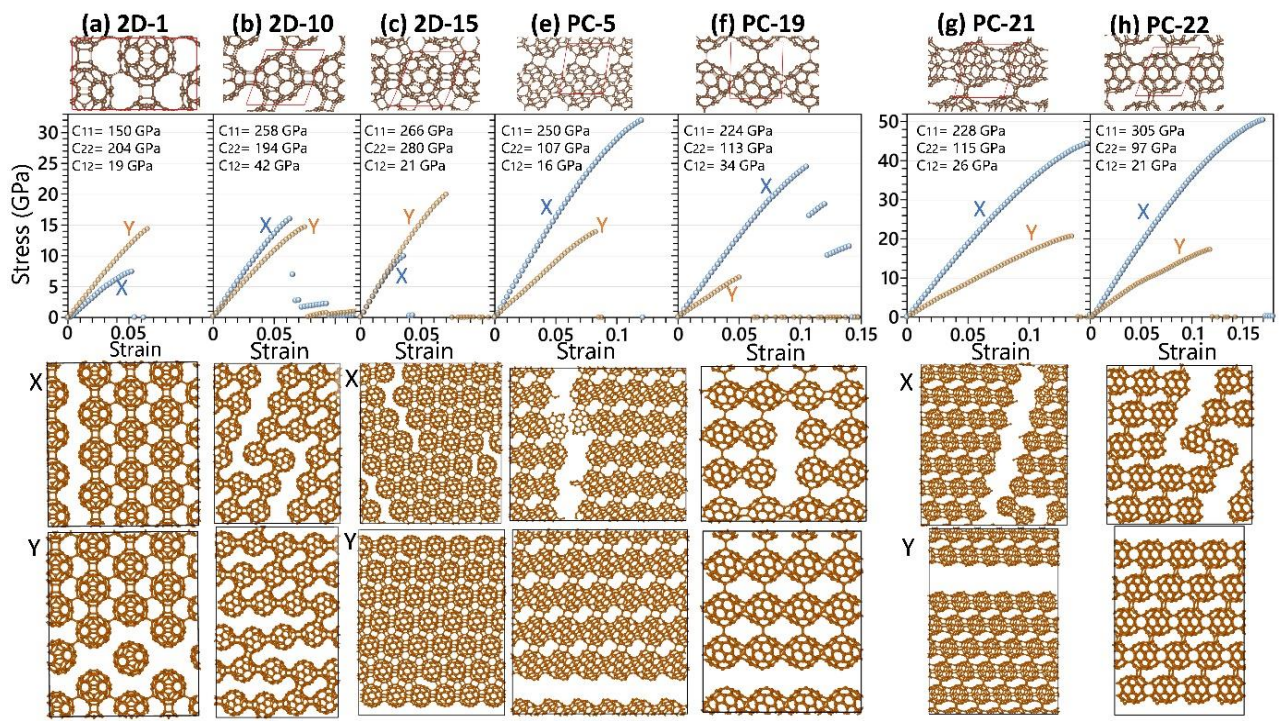

**Fig. 7**, (First row) Predicted direction dependent uniaxial stress-strain responses of the seven different $C_{60}$-based networks by MTP-based molecular dynamics simulations carried out at 1 K. X and Y directions are the horizontal and vertical directions of the presented structures, respectively. The failure mechanism for the uniaxial loading along the X and Y directions are shown beneath of the stress-strain curves in the second and third rows, respectively.

We finally employ the MTP-based molecular dynamics models to investigate the anisotropic mechanical and failure responses of the predicted $C_{60}$-based networks. Our previous investigations [38–40,45] have shown that the MTP-based simulations conducted at 1 K can accurately reproduce the ground state DFT results. Fig. 7 shows the predicted direction-dependent uniaxial stress-strain responses of the seven different $C_{60}$-based networks simulated at 1 K. In the presented results, we considered the real volumes of the deformed lattices in evaluating the stress value, assuming a fixed thickness of 8.8 Å during the uniaxial loading. The elastic constants for every lattice are also summarized in Fig. 7 results. The analysis of mechanical properties confirms anisotropic elasticity and tensile behavior in the $C_{60}$-based networks. The stress-strain curves moreover indicate that the $C_{60}$-based fullerene networks generally experience abrupt stress drops to zero value after exceeding the ultimate tensile strength point, revealing the brittle failure mechanism in these systems. An exception is however observable for the case of the PC-19 lattice, in which along the X direction the failure occurs gradually. It is noticeable that along the $C_{60}$-based crystalline 2D networks, the failure appears always along the connecting C-C bonds and the $C_{60}$ cages remain completely intact. As an interesting observation, it is conspicuous that the $C_{60}$-based porous carbons networks can stay intact at higher strains, exhibiting considerably higher elastic modulus and tensile strengths and also presenting more anisotropic behaviors than the crystalline counterparts. For example, the herein predicted PC-22 lattice, which was also found to be thermally stable at 1000 K, close to the ground state is able to yield a maximum stretchability and tensile strength values of 0.17 and 50 GPa, respectively, which are only by 40 and 58% lower than the corresponding values of the ultrahigh strong graphene [38], respectively. It is clear that porous carbons fullerene networks are extremely light-weight and low-density structures, which can exhibit remarkable mechanical characteristics.

## 4. Concluding remarks

Inspired by the latest exciting accomplishments concerning the synthesis of anisotropic and nanoporous lattices of the $C_{60}$-based fullerenes [10,19,20], in this work we employed a combination of first-principles and machine learning modeling in order to investigate the energetic stability and key physical properties of novel $C_{60}$-based fullerene lattices. We could successfully predict thermally and dynamically stable 2D, 1D, and porous-carbon $C_{60}$-based networks with close energies to the native $C_{60}$ cage. Because of the extremely close energies of the $C_{60}$-based 2D networks, in the experimentally fabricated samples various atomic configurations may co-exist simultaneously. The predicted $C_{60}$-based networks were found to possess metallic, semimetallic, or semiconducting electronic nature depending on their atomic configurations. It is shown that while the crystalline $C_{60}$-based networks mostly are semiconductors, the metallic electronic nature is more conspicuous in the porous carbons counterparts. The phonon dispersion relations, mechanical and failure responses, thermal expansion coefficients and lattice thermal conductivity were explored using accurate and computationally efficient machine learning interatomic potentials. By implementing the first-principles multiscale modelling [38,58], we could precisely reproduce the experimentally measured lattice thermal conductivity of the qHPC$_{60}$ system [20]. It is moreover concluded that the lattice thermal conductivity of the single-crystalline fullerene nanosheets at room temperature are anisotropic and generally in the order of 10 W/mK. Because of spatial atomic configurations, the formation of thermally induced out-of-plane wrinkles is very limited, resulting in the exhibition of positive thermal expansion in the $C_{60}$-based 2D networks. It is additionally found that the 2D porous-carbon $C_{60}$-based networks can exhibit exceptional mechanical properties, with tensile strengths and elastic moduli reaching values of 50 and 300 GPa, respectively. In comparison with crystalline $C_{60}$-based 2D networks, the porous carbon counterparts are found to demonstrate higher thermal stability, elastic modulus, tensile strengths and ability to remain intact under larger strains. The developed computational methodology in this study could be effectively extended to explore novel 1D and 2D networks, made of various fullerene molecules. This work moreover provides a comprehensive overview on the structural, stability, electronic, thermal, and mechanical properties of $C_{60}$-based fullerene networks.

## Acknowledgment

The author appreciates the funding by the Deutsche Forschungsgemeinschaft (DFG, German Research Foundation) under Germany's Excellence Strategy within the Cluster of Excellence PhoenixD (EXC 2122, Project ID 390833453). The author is greatly thankful to the VEGAS cluster at Bauhaus University of Weimar for providing the computational resources. The author gratefully acknowledges the computing time granted by the Resource Allocation Board and